\documentclass[aps,physrev, twocolumn, superscriptaddress]{revtex4-2}
\usepackage{amsmath}
\usepackage{graphicx}
\usepackage{bm}% bold math
\usepackage{color}
\usepackage{amsmath, amssymb,amsfonts}
\usepackage{mathrsfs}
\usepackage{type1cm}
\usepackage[colorlinks,linkcolor=blue,citecolor=blue]{hyperref}
\usepackage{physics}
\usepackage{dcolumn}
\usepackage{upgreek}
\usepackage{orcidlink}
\usepackage{qcircuit}
\usepackage{braket}
\usepackage[normalem]{ulem}

\begin{document}

%\preprint{}

\title{
Pulse-based optimization of quantum many-body states with Rydberg atoms in optical tweezer arrays
}% 

\author{Kazuma Nagao\,\orcidlink{0009-0009-0401-1958}}
\email{kazuma.nagao@riken.jp}
\affiliation{%
Computational Materials Science Research Team, RIKEN Center for Computational Science (R-CCS), Hyogo 650-0047, Japan
}%
\affiliation{%
Quantum Computational Science Research Team, RIKEN Center for Quantum Computing (RQC), Wako, Saitama 351-0198, Japan
}%
%%%
\author{Sergi Juli\`a-Farr\'e\,\orcidlink{0000-0003-4034-5786}}
\affiliation{PASQAL SAS, 24 rue Emile Baudot - 91120 Palaiseau,  Paris, France}%
\author{Joseph Vovrosh\,\orcidlink{0000-0002-1799-2830}}
\affiliation{PASQAL SAS, 24 rue Emile Baudot - 91120 Palaiseau,  Paris, France}%
%%%
\author{Alexandre Dauphin\,\orcidlink{0000-0003-4996-2561}}
\affiliation{PASQAL SAS, 24 rue Emile Baudot - 91120 Palaiseau,  Paris, France}%
%%%
\author{Seiji Yunoki\,\orcidlink{0000-0002-8167-452X}}
\affiliation{%
Computational Materials Science Research Team, RIKEN Center for Computational Science (R-CCS), Hyogo 650-0047, Japan
}%
\affiliation{%
Quantum Computational Science Research Team, RIKEN Center for Quantum Computing (RQC), Wako, Saitama 351-0198, Japan
}%
\affiliation{%
Computational Condensed Matter Physics Laboratory, RIKEN Pioneering Research Institute (PRI), Saitama 351-0198, Japan
}%
\affiliation{%
Computational Quantum Matter Research Team, RIKEN Center for Emergent Matter Science (CEMS), Saitama 351-0198, Japan
}%

\date{\today}

\begin{abstract}

We explore a pulse-based variational quantum eigensolver (VQE) algorithm for Rydberg atoms in optical tweezer arrays and evaluate its performance on prototypical quantum spin models.
We numerically demonstrate that the ground states of the one-dimensional antiferromagnetic Heisenberg model and the mixed-field Ising model can be accurately prepared using an adaptive update algorithm that randomly segments pulse sequences, for systems of up to ten qubits.
Furthermore, we propose and validate a hybrid scheme that integrates this pulse-level analog quantum algorithm with a variational quantum gate approach, where digital quantum gates are approximated by optimized analog pulses.
This enables efficient measurement of the cost function for target many-body Hamiltonians.

\end{abstract}

\maketitle

\section{
Introduction
}
\label{Sec:1}

Quantum devices composed of individually controlled quantum systems have opened new avenues for exploring fundamental problems in quantum many-body physics and quantum information science~\cite{nielsen2010quantum, browaeys2020many, foss2024progress}.  
One of the primary applications of today's commercially available quantum devices, now accessible to end users, is digital quantum computing, which utilizes single- and two-qubit gates to prepare entangled states via quantum circuits and to execute quantum algorithms.  
In recent years, digital quantum computers, such as IBM’s superconducting qubit systems and Quantinuum’s trapped-ion systems, have been successfully employed to investigate a wide range of phenomena, including discrete time crystals in Floquet circuits~\cite{mi2022time, shinjo2024unveiling}, Floquet many-body localization (MBL)~\cite{shtanko2025uncovering}, sample-based quantum diagonalization (SQD) for quantum chemistry~\cite{robledo2025chemistry}, and quantum information scrambling in chaotic many-body systems~\cite{seki2025simulating}.

Among the various quantum hardware platforms, 
neutral atoms trapped in optical tweezer arrays represent a particularly promising alternative, characterized by their inherent scalability and flexible programmability~\cite{levine2018high, scholl2021quantum}.  
Notably, neutral-atom platforms can support both analog and digital operational modes within the same architecture~\cite{Henriet_2020,lanes2025framework}. 
In the analog mode, they allow for quantum simulations of quantum spin models
by exploiting intrinsic long-range Ising~\cite{labuhn2016tunable, bernien2017probing, semeghini2021probing} and XY~\cite{scholl2022microwave, chen2023continuous} interactions mediated by Rydberg excited states, along with tunable geometries in two and three dimensions~\cite{barredo2018synthetic, scholl2021quantum}.   
Recent experimental advances include the observation of nonthermal dynamics associated with quantum many-body scar states~\cite{turner2018weak,moudgalya2022quantum,chandran2023quantum} and the realization of topological order in kagome lattice systems~\cite{semeghini2021probing}.
Moreover, the strong dipole-dipole interactions between Rydberg atoms enable the implementation of fast, high-fidelity two-qubit gates, which constitute the basic building blocks of digital quantum computing with neutral atoms~\cite{scholl2021quantum, chew2022ultrafast, bluvstein2024logical}.  
A recent notable milestone is the demonstration of 228 logical two-qubit gates across 48 logical qubits encoded in individual atoms, marking an important step toward fault-tolerant quantum computing~\cite{bluvstein2024logical}.

The programmability of Rydberg atomic systems enables the development of digital-analog hybrid algorithms~\cite{parra2020digital}, which aim to combine the respective advantages of analog operations and digital gates. 
In particular, the time-resolved controllability of pulses via coherent lasers and microwaves allows for the implementation of pulse-level variational algorithms that do not rely on explicit digital quantum gates~\cite{meitei2021gate,michel2023blueprint,sherbert2025parametrization}. 
While gate-based digital quantum algorithms are, in general, universal, their practical utility is often hindered by the accumulation of gate errors due to noise. 
The resulting degradation in state fidelity typically grows exponentially with the circuit depth and system size, placing severe limitation on the performance of quantum algorithms.
In contrast, digital-analog quantum algorithms, where the analog components are executed via direct pulse control, offer potential advantages due to the generally higher fidelities of analog operations compared to digital gates~\cite{meitei2021gate, michel2023blueprint, andersen2025thermalization}. Hence, they are increasingly regarded as a promising alternative pathway toward universal quantum computation~\cite{parra2020digital}.

Among the various algorithms proposed, pulse-based variational quantum eigensolver (VQE) methods~\cite{meitei2021gate,asthana2023leakage,michel2023blueprint,sherbert2025parametrization,shingu2025digital} have gained significant attention. 
Applications in quantum chemistry have been explored in the contexts of nonlinear plasmon systems~\cite{meitei2021gate,sherbert2025parametrization} and Rydberg atoms~\cite{michel2023blueprint}. 
However, since analog operations are generally not universal, performing general-purpose tasks, such as experimentally evaluating cost functions for arbitrary Hamiltonians, may require the incorporation of digital quantum gates as auxiliary tools.
The available {\it measurement toolbox} for estimating expectation values of generic Hamiltonians remains limited, although recent advances include randomized measurement protocols~\cite{elben2023randomized,notarnicola2023randomized}, derandomization techniques for Pauli string evaluation~\cite{huang2021efficient, michel2023blueprint}, and classical shadow tomography methods~\cite{shingu2025digital}.  
Moreover, most exciting applications of pulse-based VQE have focused on quantum chemistry, despite the method's natural applicability to a broader class of systems, such as those in condensed matter physics and quantum magnetism.

In this paper, 
we investigate a pulse-based VQE algorithm implemented on Rydberg-atom analog quantum simulators and evaluate its performance on prototypical quantum spin models. 
We demonstrate that the ground states of the one-dimensional Heisenberg and mixed-field Ising models can be accurately prepared using an adaptive pulse segmentation scheme 
inspired by a variant of the ctrl-VQE approach~\cite{meitei2021gate, michel2023blueprint, parametrization2025sherbert}, in which the number of variational parameters increases with each iterative subdivision of the global pulse sequence. 
To facilitate realistic measurements of target Hamiltonians, we further propose the incorporation of a variational quantum gate strategy developed in Ref.~\cite{chevallier2024variational}, where both global and local quantum gates are approximately synthesized via pulse-level optimization.

The rest of this paper is organized as follows. 
In Sec.~\ref{Sec:2}, we review the fundamental aspects of Rydberg atoms in optical tweezer arrays. 
Section~\ref{Sec:3} introduces the pulse-based VQE algorithm tailored for Rydberg quantum simulators with Ising-type interactions. 
In Sec.~\ref{Sec:4}, we present our main numerical results for the one-dimensional antiferromagnetic Heisenberg model, 
demonstrating that the pulse-based VQE ansatz can accurately reproduce its exact ground state for systems up to ten qubits. 
Section~\ref{Sec:5} provides results for the one-dimensional mixed-field Ising model, whose ground state is shown to be more efficiently described than in the Heisenberg case. 
In Sec.~\ref{Sec:6}, we discuss a hybrid approach that integrates the pulse-based VQE with variational quantum gate synthesis. 
Finally, Sec.~\ref{Sec:7} summarizes our findings and outlines future research directions. 
Additional numerical details are provided in Appendix~\ref{app:iterations}, and a protocol for preparing the initial state with momentum $q=\pi$ is given in Appendix~\ref{app:state_preparation}.

%%%%%%%%%%%%%%%%%%%%%%%%%%%%%%%%%%%%%%%%%%
%%%%%%%%%%%%%%%%%%%%%%%%%%%%%%%%%%%%%%%%%%
\section{
Rydberg atoms in optical tweezer arrays
}
\label{Sec:2}
%%%%%%%%%%%%%%%%%%%%%%%%%%%%%%%%%%%%%%%%%%
%%%%%%%%%%%%%%%%%%%%%%%%%%%%%%%%%%%%%%%%%%

Rydberg atoms in optical tweezer arrays offer a versatile and programmable platform for realizing highly controlled pseudospin-$1/2$ systems~\cite{browaeys2020many}.
Specifically, we focus on $N$ rubidium-87 ($^{87}$Rb) atoms in an array~\cite{levine2018high, scholl2021quantum}, where the ground state of an atom ($\ket{g} = \ket{\downarrow}$) and a Rydberg excited state ($\ket{r}=\ket{\uparrow}$) form an effective two-level system representing a qubit.
Due to the strong dipole-dipole interactions between atoms in the Rydberg state, 
the system is described by a pseudospin-$1/2$ Hamiltonian with long-range antiferromagnetic Ising interactions~\cite{browaeys2020many}
\begin{align}\label{eq:IsingHamiltonian}
    {\hat H}(t)
    &= \sum_{i < j} J_{ij} {\hat n}_{i} {\hat n}_{j} - \hbar \Delta(t) \sum_{j=1}^{N} {\hat n}_j  \\
    &\;\;\;\;\;\;\;\; + \frac{\hbar \Omega(t)}{2}  \sum_{j=1}^{N} \left( \ket{g} \bra{r}_j +  \ket{r} \bra{g}_j  \right), \nonumber
\end{align}
where ${\hat n}_j = \ket{r}\bra{r}_j$ is the number operator for Rydberg excitation at the $j$th atom and $\hbar = h/2\pi$ is the reduced Planck constant. 
The potential energy function $J_{ij}$ describes the van der Waals interactions between atoms $i$ and $j$, and decays with distance as $J_{ij} = C_6/r_{ij}^6$, where $C_6$ is a constant.
In typical experimental configurations, optical tweezer arrays are arranged in an arbitrary two-dimensional geometry $G$, and the interatomic distance is given by $r_{ij} = || {\vec R}_{i} - {\vec R}_{j} || = \sqrt{(x_i - x_j)^2 + (y_i - y_j)^2}$, where ${\vec R}_{i},{\vec R}_{j} \in G$ denote the atomic positions. 
The first term in Eq.~(\ref{eq:IsingHamiltonian}) contributes only when both atoms $i$ and $j$ are simultaneously excited to the Rydberg state, giving rise to the Rydberg blockade effect~\cite{browaeys2020many}. 
The time-dependent functions $\Delta(t)$ and $\Omega(t)$ represent the laser detuning and Rabi frequency, respectively. 
For simplicity, we neglect the phase of the Rabi frequency and treat $\Omega(t)$ as a real-valued function. 
Throughout this work, we focus on global pulse protocols; hence, both $\Delta(t)$ and $\Omega(t)$ are assumed to be spatially uniform.

For later convenience, we define the local Pauli operators associated with the atomic ground and Rydberg states as 
\begin{align}\label{eq:Paulis}
{\hat X}_{j} &= \ket{g}\bra{r}_j + \ket{r}\bra{g}_j, \nonumber \\
{\hat Y}_{j} &= i \ket{g}\bra{r}_j - i \ket{r}\bra{g}_j, \\
{\hat Z}_{j} &= \ket{r}\bra{r}_j - \ket{g}\bra{g}_j. \nonumber
\end{align}
These operators are Hermitian and satisfy the standard Pauli commutation relations, such as $[{\hat X}_{j},{\hat Y}_{j}] = 2 i {\hat Z}_{j}$.
With this notation, the detuning and Rabi-coupling terms in Eq.~(\ref{eq:IsingHamiltonian}) can be rewritten as $-\frac{\Delta(t)}{2}\sum_{j}{\hat Z}_j - \frac{N\Delta(t)}{2}$ and $\frac{\Omega(t)}{2}\sum_{j}{\hat X}_{j}$, respectively. Thus, these terms correspond to the longitudinal and transverse field couplings in quantum magnets.

%%%%%%%%%%%%%%%%%%%%%%%%%%%%%%%%%%%%%%%%%%%%%%%%%%%%%%%%%%%
%%%%%%%%%%%%%%%%%%%%%%%%%%%%%%%%%%%%%%%%%%%%%%%%%%%%%%%%%%%
\section{
Pulse-based variational quantum eigensolver 
}
\label{Sec:3}
%%%%%%%%%%%%%%%%%%%%%%%%%%%%%%%%%%%%%%%%%%%%%%%%%%%%%%%%%%%
%%%%%%%%%%%%%%%%%%%%%%%%%%%%%%%%%%%%%%%%%%%%%%%%%%%%%%%%%%%

\begin{figure} 
\includegraphics[width=\columnwidth]{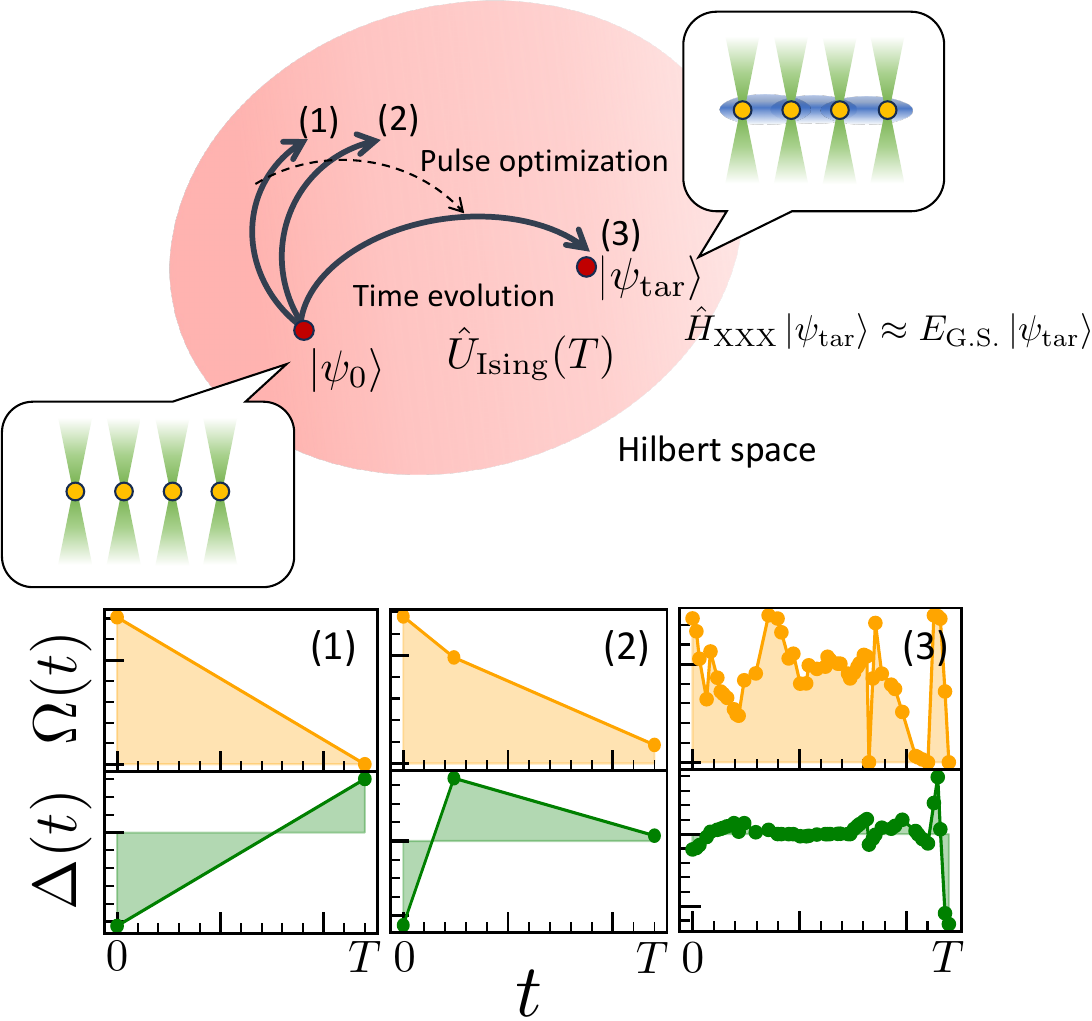}
\vspace{0mm}
\caption{
Schematic illustration of the pulse-based variational quantum eigensolver (PVQE) for neutral atoms in an optical tweezer array.
(Top) 
Pulse level optimization of the unitary operator ${\hat U}_{\rm Ising}(T) = {\cal T}_{t}\exp{-\frac{i}{\hbar}\int^{T}_{0} dt {\hat H}(t)}$ in the analog Rydberg-atom system. 
A given instance of time-dependent control parameters--Rabi frequency $\Omega(t)$ and detuning $\Delta(t)$--along with time-independent system parameters, defines a trajectory of quantum dynamics starting from the initial state $\ket{\psi_0}$ in the Hilbert space. 
The evolution is governed by the unitary operator ${\hat U}_{\rm Ising}(T)$ for the Ising Hamiltonian defined in Eq.~(\ref{eq:IsingHamiltonian}).
The flowchart labeled (1), (2), and (3) illustrates how an initial pulse sequence is progressively refined to minimize a cost function associated with a target Hamiltonian, such as the Heisenberg model. 
Once optimized, the analog evolution may approximate the target state, i.e., $\ket{\psi_{\rm tar}} \approx {\hat U}_{\rm Ising}(T)\ket{\psi_0}$. 
(Bottom) 
Examples of pulse optimization through random time segmentation. 
Panel (1) shows the initial single-segment pulse profile for $\Omega(t)$ and $\Delta(t)$. 
Panel (2) illustrates a two-segmentation profile obtained by splitting the previous pulse sequence at a randomly selected time point. 
Panel (3) depicts a refined sequence after multiple rounds of segmentation and classical optimization at the segmentation boundaries. 
}
\label{fig1}
\end{figure}

The pulse-based VQE, hereafter referred to as PVQE, utilizes the controllability of pulse sequences in a physical system~\cite{meitei2021gate,michel2023blueprint,sherbert2025parametrization}. 
In the Rydberg-atom system described by Eq.~(\ref{eq:IsingHamiltonian}), the Rabi frequency $\Omega(t)$ and the detuning $\Delta(t)$ are time-dependent control parameters of the driving laser fields. 
In addition, the spatial configuration of the optical tweezer arrays can be tuned during the qubit initialization stage, thereby determining the geometry-dependent Ising interaction energy $J_{ij}$ in Eq.~(\ref{eq:IsingHamiltonian}).
Given these parameters, we define a variational state vector for a target quantum state as 
\begin{align}\label{eq:pvqe_ansatz}
    \ket{\Psi[\Omega,\Delta,G]} = {\cal T}_{t}\exp {-\frac{i}{\hbar}\int^{T}_{0}dt {\hat H}(t) }\ket{\psi_0},
\end{align}
where ${\cal T}_{t}$ denotes the time-ordering operator (from right to left), $T$ is the total pulse duration, and $\ket{\psi_0}$ is a reference state. 
By varying the control parameters, $\Omega(t)$, $\Delta(t)$, and $G$, to minimize a cost function, one generates a family of quantum trajectories originating from $\ket{\psi_0}$ in the Hilbert space of the Rydberg atoms, as illustrated in Fig.~\ref{fig1}. 
In this work, we treat the total pulse duration $T$ as a fixed parameter during the variational optimization. 
Furthermore, we assume that $G$ belongs to a family of geometries characterized by a single real parameter.

%%%%%%%%%%%%%%%%%%%%
%%%%%%%%%%%%%%%%%%%%
%%%%%%%%%%%%%%%%%%%%
%%%%%%%%%%%%%%%%%%%%

To optimize the pulse profiles $\Omega(t)$ and $\Delta(t)$, we employ a generalized adaptive update algorithm~\cite{meitei2021gate, michel2023blueprint}, in which the pulse sequences are randomly and iteratively segmented. 
As a first step, we initialize $N$ qubits in the desired geometry $G$ using the atom rearrangement technique for optical tweezers~\cite{scholl2021quantum}. 
We then apply a linearly scheduled pulse sequence of the form 
\begin{align}
\Omega(t) 
&= \Omega_{0}\left(1- \frac{t}{T}  \right) + \Omega_{1}\frac{t}{T}, \\
%%%
\Delta(t) 
&= \Delta_{0}\left(1- \frac{t}{T}  \right) + \Delta_{1}\frac{t}{T},
\end{align}
where $(\Omega_0, \Omega_1, \Delta_0, \Delta_1)$ are the variational parameters, corresponding to the pulse amplitudes at the start and end of the time window $[0,T]$. 
These parameters are optimized to minimize the expectation value of a target Hamiltonian ${\hat H}_{\rm tar}$, 
\begin{align}
    E_{\rm cost}[\Omega,\Delta,G] = \bra{\Psi[\Omega,\Delta,G]} {\hat H}_{\rm tar} \ket{\Psi[\Omega,\Delta,G]}.
\end{align}
Starting from an initial set of parameters $(\Omega_0, \Omega_1, \Delta_0, \Delta_1, G)$, we perform a classical variational optimization and obtain an updated set ${\vec \Theta}^*_{1} =(\Omega_0^*, \Omega_1^*, \Delta_0^*, \Delta_1^*, G^*)$.
This completes the first iteration of the PVQE procedure.

To initiate the second optimization step, we randomly split the trained pulse sequence characterized by $(\Omega_0^*, \Omega_1^*, \Delta_0^*, \Delta_1^*)$ into two segments. 
This segmentation divides the original time interval $[t_0,t_1] \equiv [0,T]$ into $[t_0,t_{ s},t_1]$, where $t_s \in (0,T)$ is a randomly selected time point. 
We note that both subintervals must satisfy a minimum duration constraint: $t_s - t_0>\tau_d$ and $t_1 - t_s>\tau_d$, where $\tau_d$ is chosen to be $16$~ns. 
Furthermore, $t_s$ must be an integer multiple of the system clock period, which is $4$~ns~\cite{silverio2022pulser}.
This lower bound is necessary because physical hardware cannot accommodate arbitrarily fast changes in the laser fields. 
We set the values of $\Omega(t)$ and $\Delta(t)$ at $t=t_s$ by linear interpolation as   
\begin{align}
    \begin{bmatrix}
        \Omega_{s} \\
        \Delta_{s}
    \end{bmatrix}
    = \frac{(t_1 - t_s)}{t_1 - t_{0}}
    \begin{bmatrix}
        \Omega^*_{0} \\
        \Delta^*_{0}
    \end{bmatrix}
     + \frac{(t_s - t_{0})}{t_1 - t_{0}}
    \begin{bmatrix}
        \Omega^*_{1} \\
        \Delta^*_{1}
    \end{bmatrix}.
\end{align}
We then redefine the variational parameters as $(\Omega^*_0,\Omega_s,\Omega^*_1, \Delta^*_0,\Delta_s,\Delta^*_1,G^*) \rightarrow (\Omega_0,\Omega_1,\Omega_2,\Delta_0,\Delta_1,\Delta_2,G)$ and the time intervals as $[t_0,t_s,t_1] \rightarrow [t_0,t_1,t_2]$.
At this level of approximation, the state-vector ansatz $\ket{\Psi[\Omega,\Delta,G]}$ now depends on seven variational parameters. 
A second round of classical optimization is then performed, yielding an update parameter set ${\vec \Theta}^*_{2} = (\Omega^*_0,\Omega^*_1,\Omega^*_2,\Delta^*_0,\Delta^*_1,\Delta^*_2,G^*)$.

We repeat the segmentation and optimization procedure described above until the desired level of accuracy is achieved. 
At the $M$th iteration step, one of the existing time intervals $[t_{i-1}, t_i]$ is randomly selected and further subdivided as $[t_0,\cdots,t_{i-1},t_i,\cdots,t_{M-1}] \rightarrow [t_0,\cdots,t_{i-1},t_s,t_{i},\cdots,t_{M-1}]$, where $t_s\in(t_{i-1},t_i)$ denotes the new splitting point. The values of $\Omega(t)$ and $\Delta(t)$ at $t=t_s$ are determined by linear interpolation as 
\begin{align}
    \begin{bmatrix}
        \Omega_{s} \\
        \Delta_{s}
    \end{bmatrix}
    = \frac{(t_i - t_s)}{t_i - t_{i-1}}
    \begin{bmatrix}
        \Omega^*_{i-1} \\
        \Delta^*_{i-1}
    \end{bmatrix}
     + \frac{(t_s - t_{i-1})}{t_i - t_{i-1}}
    \begin{bmatrix}
        \Omega^*_{i} \\
        \Delta^*_{i}
    \end{bmatrix}.
\end{align}
The total number of variational parameters is now $2M+3$, and the classical optimization is performed to obtain the optimized parameter set 
\begin{align}
{\vec \Theta^*_{M}} = (\Omega^*_0,\cdots,\Omega^*_M,\Delta^*_0,\cdots,\Delta^*_M,G^*).
\end{align}
Finally, the optimized unitary operator, representing the time-ordered evolution under the time-dependent Hamiltonian $\hat H(t)$ in Eq.~(\ref{eq:IsingHamiltonian}), is approximated as 
\begin{align}
{\hat U}({\vec \Theta^*_{M}}) = \hat U(t_M,t_{M-1}) \cdots \hat U(t_1,t_0),
\label{eq:Uising}
\end{align}
where each segment is defined by 
\begin{align}
\hat U(t_{i},t_{i-1})={\cal T}_{t}\exp {-\frac{i}{\hbar}\int^{t_i}_{t_{i-1}}dt {\hat H}(t) }
\end{align}
with linearly interpolated pulse profiles 
\begin{align}
\Omega(t)=\frac{\Omega_i^*-\Omega_{i-1}^*}{t_i-t_{i-1}}(t-t_{i-1}) + \Omega_{i-1}^*, \\
\Delta(t)=\frac{\Delta_i^*-\Delta_{i-1}^*}{t_i-t_{i-1}}(t-t_{i-1}) + \Delta_{i-1}^*.
\end{align}

It is worth noting that typical implementations of PVQE~\cite{meitei2021gate,michel2023blueprint} employ piecewise-constant pulses, which involve abrupt rises and falls, and are therefore not directly implementable on real quantum devices.
In contrast, our scheme, based on linearly varying pulses, is more experimentally realistic, as it incorporates a lower bound on the segmentation duration $t_{i+1}-t_{i}$ imposed by hardware limitations, as discussed earlier. 
However, when the pulse segmentation becomes sufficiently fine, the optimized pulse profiles tend to exhibit sharp spikes, as illustrated in Fig.~\ref{fig3}(b) in the next section. 
Such rapid variations are challenging to faithfully reproduce on actual hardware. 
This issue could potentially be addressed by employing differentiable pulse modulation strategies, such as those offered by the recently developed extension package PulserDiff~\cite{abramavicius2025pulserdiff}, although a detailed investigation of this approach is beyond the scope of the present work.

%%%%%%%%%%%%%%%%%%%%%%%%%%%%%%%%%%%%%%%%%%%%%%%%%%%%%%%%
%%%%%%%%%%%%%%%%%%%%%%%%%%%%%%%%%%%%%%%%%%%%%%%%%%%%%%%%
\section{
PVQE for the Heisenberg model 
}
\label{Sec:4}
%%%%%%%%%%%%%%%%%%%%%%%%%%%%%%%%%%%%%%%%%%%%%%%%%%%%%%%%
%%%%%%%%%%%%%%%%%%%%%%%%%%%%%%%%%%%%%%%%%%%%%%%%%%%%%%%%

In this section, we apply the PVQE method described in Sec.~\ref{Sec:3} to the one-dimensional antiferromagnetic Heisenberg model, also known as the XXX model~\cite{takahashi1999thermodynamics}. 
This model describes nearest-neighbor antiferromagnetic spin-exchange interactions with full SU(2) symmetry, and its Hamiltonian is given by 
\begin{align}\label{eq: AFH}
{\hat H}_{\rm XXX}
= J \sum_{i=1}^{N}{\hat {\bm S}}_{i} \cdot {\hat {\bm S}}_{i+1},
\end{align}
where $J>0$ is the coupling strength, and $\hat{\bm S}_{i} = ({\hat S}^{x}_{i},{\hat S}^{y}_{i},{\hat S}^{z}_{i}) = \frac{1}{2}({\hat X}_{i},{\hat Y}_{i},{\hat Z}_{i})$ denotes the spin-1/2 operator at site $i$.
We impose periodic boundary conditions such that ${\hat {\bm S}}_{N+1}={\hat {\bm S}}_{1}$. 
It is well known that the Hamiltonian in Eq.~(\ref{eq: AFH}) is exactly solvable analytically via the Bethe ansatz~\cite{takahashi1999thermodynamics,faddeev1996algebraic}. 
Circuit-based VQE studies on the Heisenberg model and its extensions can be found in Refs.~\cite{seki2020symmetry,jattana2022assessment,bosse2022probing,fujii2022deep,RYSun2023}.

The target state of interest is the ground state $\ket{\Psi_{\rm G.S.}}$ of the Hamiltonian ${\hat H}_{\rm XXX}$, which satisfies ${\hat H}_{\rm XXX} \ket{\Psi_{\rm G.S.}} = E_{\rm G.S.} \ket{\Psi_{\rm G.S.}}$.
Since the target system assumes periodic boundary conditions, it is natural to arrange the qubits in a circular geometry of radius $R$. 
This configuration exhibits discrete rotational symmetry, with qubit positions invariant under rotations by integer multiples of the unit angle ${\overline \varphi} = 2\pi/N$.
The cost function to be minimized is given by $E_{\rm cost}^{\rm XXX}[\Omega,\Delta,R] \equiv \bra{\Psi [\Omega,\Delta,R]}{\hat H}_{\rm XXX} \ket{\Psi[\Omega,\Delta,R]}$.
The reference state is chosen as the fully polarized product state $\ket{\psi_0} = \bigotimes_{j=1}^{N}\ket{g}_j = \bigotimes_{j=1}^{N}\ket{ \downarrow }_j$, which is a typical initial state prepared during the atom rearrangement stage in optical tweezer experiments.  
We set the total pulse duration to $T = 2.4 \; \upmu \text{s}$, which is well below the typical coherence time of the system. 
For the Rydberg state $\ket{r}$, we use the principal quantum number $n=70$ of $^{87}\text{Rb}$, for which the interaction strength is $C_6/\hbar = 5420158.53$ rad $\upmu \text{m}^6$ $\upmu\text{s}^{-1}$~\cite{silverio2022pulser}.
Classical optimization is performed using the Nelder-Mead algorithm implemented in SciPy~\cite{2020SciPy-NMeth,nelder1965simplex}, 
which conducts gradient-free parameter searches to minimize the cost function.
The parameters $\Omega(t)$ and $\Delta(t)$ are constrained within experimentally accessible bounds: $\Omega(t) \in [0,15]$ $\text{rad}\; \upmu\text{s}^{-1}$ and $\Delta(t) \in [-125,125]$ $\text{rad}\; \upmu\text{s}^{-1}$~\cite{silverio2022pulser}.
The maximum number of optimization iterations for the Nelder-Mead algorithm is set to 5000. 
All numerical simulations in this work are carried out using the Pulser library~\cite{silverio2022pulser}, which provides virtual simulators closely aligned with the capabilities of real neutral-atom quantum devices.

%%%%%%%%%%%%%%%%%%%%
%%%%%%%%%%%%%%%%%%%%
\begin{figure*} 
\includegraphics[width=\textwidth]{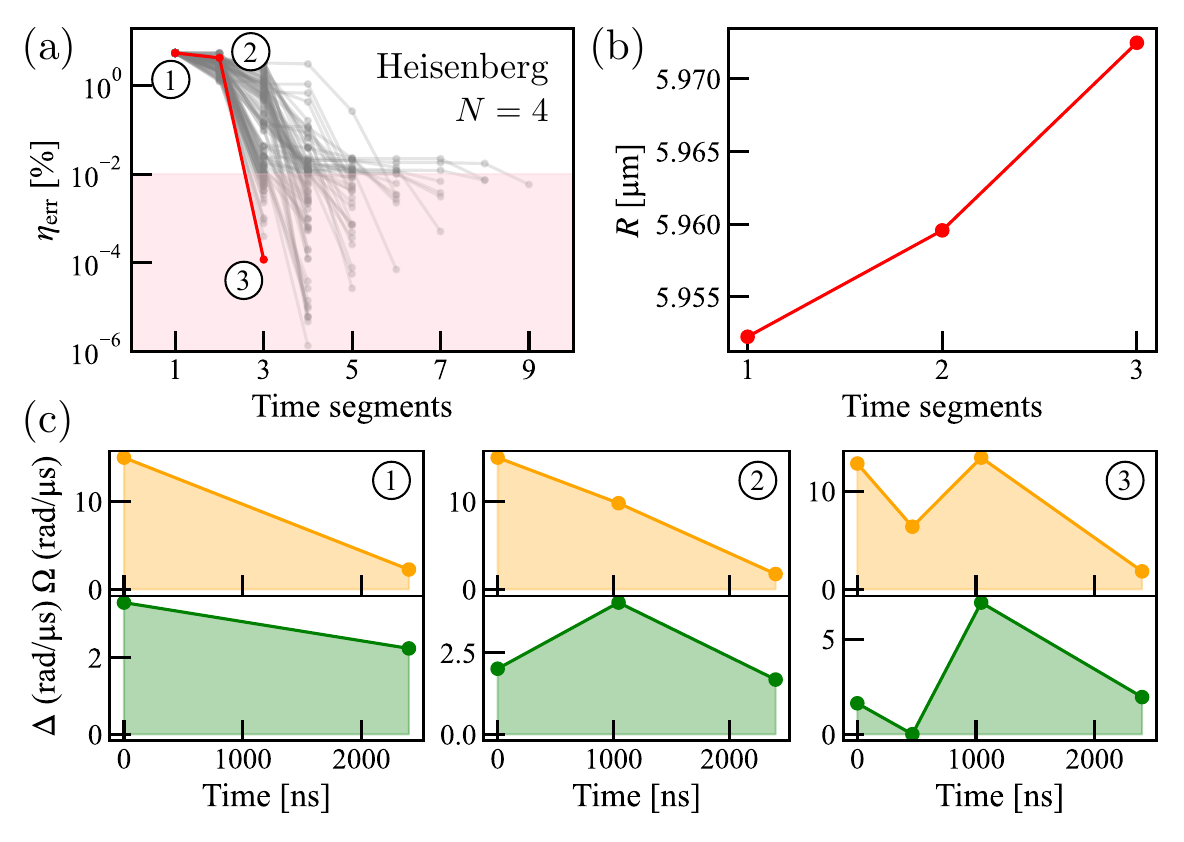}
\vspace{-6mm}
\caption{
Pulse-based VQE for the one-dimensional antiferromagnetic Heisenberg model with $N=4$ under periodic boundary conditions. 
(a)
Relative error $\eta_{\rm err}$ tracked at each step of the time splitting process for 100 independent PVQE optimization runs. 
The total pulse duration is fixed at $T=2.4$ $\upmu\text{s}$. 
The pink-shaded region indicates the target threshold $\epsilon = 0.01$: once  
$\eta_{\rm err}$ falls below this value, the time-splitting loop is terminated. 
The red line with circles highlights the best-performing run with the minimum number of segments. 
(b) 
Optimized radius $R$ as a function of the number of time segments, corresponding to the best run in panel (a). 
(c)
Optimized pulse profiles for the Rabi frequency $\Omega(t)$ and detuning $\Delta(t)$ at each step of time splitting, corresponding to the best run in panel (a). 
}
\label{fig2}
\end{figure*}
%%%%%%%%%%%%%%%%%%%%
%%%%%%%%%%%%%%%%%%%%

We begin by presenting results for even values of $N/2$, obtained using 
the random time-splitting algorithm described in Sec.~\ref{Sec:3}.
Figure~\ref{fig2}(a) shows 100 independent optimization runs for $N=4$, where we track the relative error defined as $\eta_{\rm err} \equiv 100\times |E^{\rm XXX}_{\rm cost} - E_{\rm G.S.}|/|E_{\rm G.S.}|$. 
Overall, the optimization based on time segmentation proves to be highly efficient, achieving high accuracy with as few as nine time segments.
For instance, the run highlighted in red in Fig.~\ref{fig2}(a) achieves a relative error $\eta_{\rm err}<\epsilon = 0.01$ using only three pulse segments. 
The evolution of the corresponding pulse shapes across optimization steps is shown in Fig.~\ref{fig2}(c). 
In the third iteration of this sequence, the optimized Rabi frequency $\Omega(t)$ and detuning $\Delta(t)$, obtained via the Nelder-Mead algorithm, exhibit four pulse edges, with the shortest segment duration around 500~ns. 
This result demonstrates the experimental feasibility of the time-splitting PVQE scheme, wherein the exact ground state of the four-site Heisenberg model is accurately prepared. 
Figure~\ref{fig2}(b) displays the evolution of the optimized radius $R$ as a function of the number of time segments, corresponding to the red trace in Fig.~\ref{fig2}(a). 
For a single pulse segment, the optimized radius is $R = 5.952 \; \upmu\text{m}$, corresponding to a nearest-neighbor Ising interaction $J_{\rm nn}/h \equiv C_6/(h d^6_{\rm nn}) \sim 2.42$ MHz, where $d_{\rm nn} = 2R \sin (\pi/N)$ denotes the nearest-neighbor distance. 
When the number of segments increases to three, the optimized radius slightly increases to $R = 5.973 \; \upmu\text{m}$, yielding $J_{\rm nn}/h = 2.38$ MHz. 
Note, however, that this variation is smaller than the typical static positional disorder in experiments, estimated as $\delta R\approx 0.1\, \mu\text{m}$~\cite{scholl2021quantum}.
In the best-performing run, a total of $6935$ Nelder-Mead iterations were performed during the final step of the time-splitting procedure (see Appendix~\ref{app:iterations} for details).

%%%%%%%%%%%%%%%%%%%%
%%%%%%%%%%%%%%%%%%%%
\begin{figure*} 
\includegraphics[width=\textwidth]{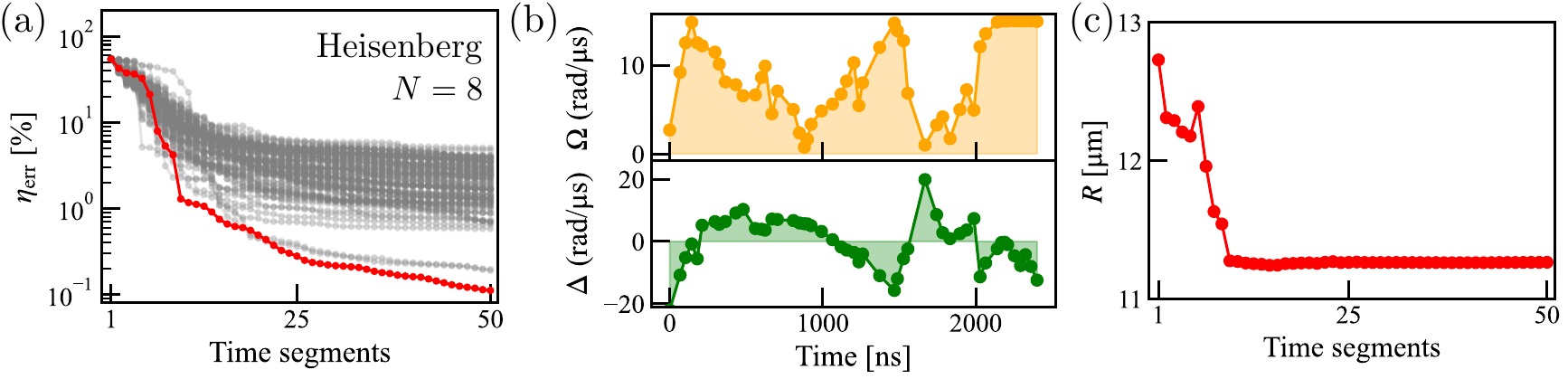}
\vspace{-6mm}
\caption{
Pulse-based VQE for the one-dimensional antiferromagnetic Heisenberg model with $N=8$ under periodic boundary conditions. 
(a)
Relative error $\eta_{\rm err}$ tracked at each step of the time-splitting process for 100 independent PVQE optimization runs. 
The total pulse duration is fixed at $T=2.4$ $\upmu\text{s}$. 
The red line with circles highlight the best-performing run, which achieves the smallest final relative error. 
(b)
Optimized pulse profiles for the Rabi frequency $\Omega(t)$ and detuning $\Delta(t)$ at the final iteration, consisting of fifty linearly varying segments. 
These results correspond to the best run in panel (a). 
(c)
Optimized radius $R$ as a function of the number of time segments, corresponding to the best run in panel (a). 
The radius converges to a stable value, yielding a nearest-neighbor Ising interaction of $J_{\rm nn}/h \approx 2.1$ MHz. 
}
\label{fig3}
\end{figure*}
%%%%%%%%%%%%%%%%%%%%
%%%%%%%%%%%%%%%%%%%%

Figure~\ref{fig3}(a) shows the relative error $\eta_{\rm err}$ as a function of the number of time segments, obtained from 100 independent PVQE optimization runs for $N=8$. As indicated by the red line with circles, the best-performing run achieves a relative error $\eta_{\rm err}$ on the order of $10^{-1}$\%, demonstrating that the PVQE remains capable of high accuracy even in larger systems. 
However, convergence is less efficient than in the $N=4$ case, requiring as many as 50 time-splitting cycles or even more. 
In this sample set, over 90 of the 100 realizations converge to states with rather higher errors. 
The ensemble average of the final relative error $\eta_{\rm err}$ at 50 time segments and its standard deviation are estimated to be $2.14 \pm 0.12$\%.
This reduced efficiency can be attributed to the two main factors: the stronger quantum correlations present in the $N=8$ ground state and the absence of explicit SU(2) symmetry in the variational ansatz defined in Eq.~(\ref{eq:pvqe_ansatz}). 
These challenges are less severe in the $N=4$ case, where the system has fewer spins and a larger finite-size gap. 
Furthermore, we estimate the ensemble-averaged final radius $\langle R\rangle =12.11 \pm 0.03\; \upmu \text{m}$, corresponding to a nearest-neighbor Ising interaction of $\langle  J_{\rm nn} \rangle / h \approx 1.36$ MHz. 
By contrast, 
the best-performing run yields a smaller final radius of $R = 11.27 \; \upmu \text{m}$, implying $J_{\rm nn}/h \approx 2.1$ MHz.
The evolution of the optimized radius $R$ as a function of the number of time segments for the best run is shown in Fig.~\ref{fig3}(c), where the values stabilize for segment counts exceeding ten. 
However, based on our numerical results, we heuristically conclude that large values of $R$ are neither sufficient nor necessary to construct an efficient variational path. 
Figure~\ref{fig3}(b) shows the optimized pulse sequence that leads to the best results. Both $\Omega(t)$ and $\Delta(t)$ consist of 50 linearly varying segments, each defined over a distinct time interval. 
The total number of Nelder-Mead iterations required at the final step of time splitting is 236\;495.

%%%%%%%%%%%%%%%%%%%%
%%%%%%%%%%%%%%%%%%%%
\begin{figure} 
\includegraphics[width=\columnwidth]{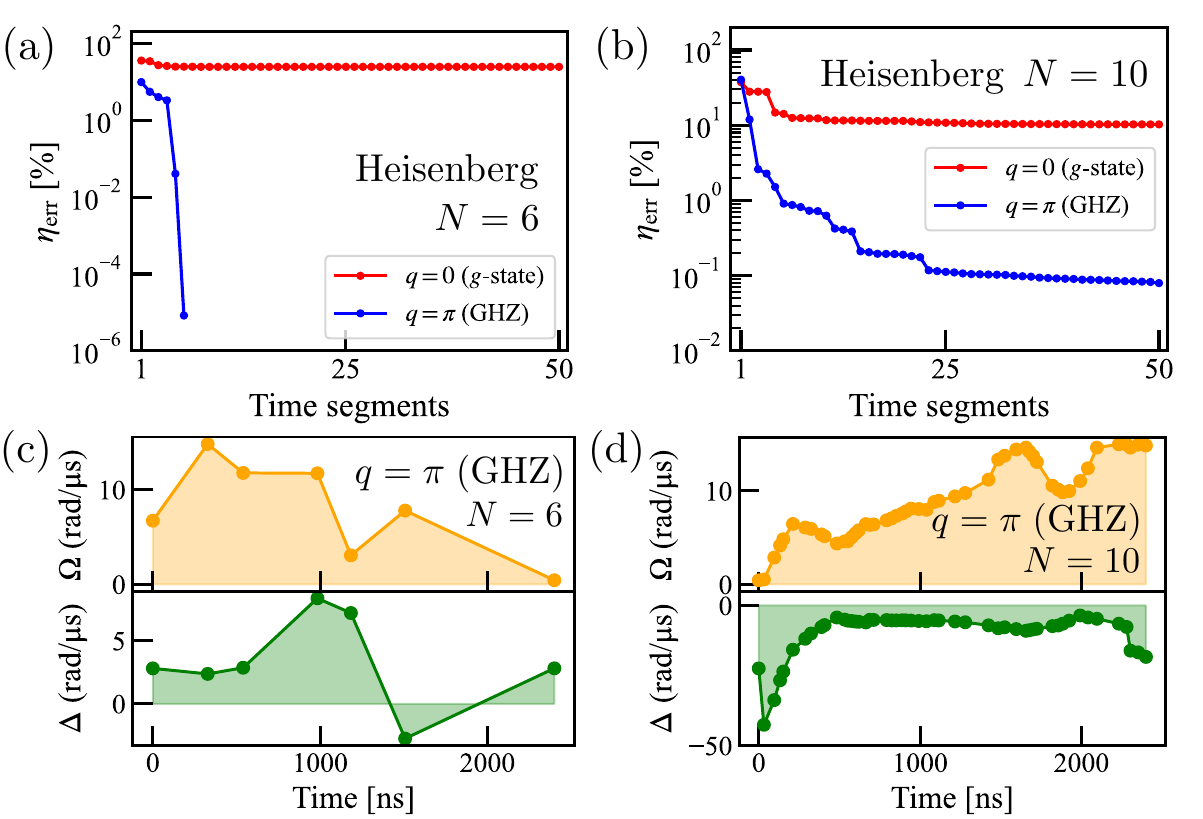}
\vspace{-6mm}
\caption{
Pulse-based VQE for the one-dimensional antiferromagnetic Heisenberg model with odd $N/2$ under periodic boundary conditions. 
(a) 
Relative error $\eta_{\rm err}$ tracked at each step of the time-splitting process for $N=6$, using two distinct initial states. 
The red line with circles shows a representative run initialized from the product $g$-state, $\ket{g \cdots g}$, which fails to converge to the target ground state and becomes trapped in a spurious plateau.  
The blue line with circles shows the best-performing result among 100 independent optimization runs initialized from the $q=\pi$ state, achieving a relative error $\eta_{\rm err}$ below the threshold $\epsilon = 0.01$ with six time segments.  
(b) 
Same as panel (a), but for $N=10$.  
The best result among 92 independent optimization runs initialized from the $q=\pi$ state achieves a relative error $\eta_{\rm err}$ on the order of $10^{-1}$\%.  
(c) 
Optimized pulse profiles for the Rabi frequency $\Omega(t)$ and detuning $\Delta(t)$ at the final iteration, consisting of six linearly varying segments. These results correspond to the best-performing run with the $q=\pi$ initial state for $N=6$ in panel (a). 
(d) 
Same as panel (c), but for $N=10$, with fifty linearly varying segments.
In all simulations, the total pulse duration is fixed at $T=2.4$ $\upmu\text{s}$. 
}
\label{fig4}
\end{figure}
%%%%%%%%%%%%%%%%%%%%
%%%%%%%%%%%%%%%%%%%%

Next, we turn to the case where $N/2$ is odd. 
In this setting, we numerically observe that any optimization process initialized from the product $g$-state, i.e., a product of ground-state atomic configurations $\ket{\psi_0}=\bigotimes_{j=1}^{N} \ket{g}_j$, fails to converge to the target ground state and becomes trapped in a local minimum.
Representative results for $N=6$ and $N=10$ under periodic boundary conditions are shown in Figs.~\ref{fig4}(a) and \ref{fig4}(b), respectively, where the relative error $\eta_{\rm err}$ saturates at a plateau on the order of $10$\%.
This failure can be attributed to the Marshall sign rule for the ground state of the Heisenberg model~\cite{marshall1955antiferromagnetism,seki2020symmetry}. 
According to this rule, the ground state is an eigenstate of the single-site translation operator ${\hat T}$ with momentum $q=\pi$ when $N/2$ is odd, and $q=0$ when $N/2$ is even. 
The operator ${\hat T}$ performs a cyclic bit-shift operation; for example, for six qubits, ${\hat T}\ket{a}_{1}\ket{b}_{2}\ket{c}_{3} \ket{d}_{4} \ket{e}_{5} \ket{f}_6 = \ket{f}_{1} \ket{a}_{2}\ket{b}_{3}\ket{c}_{4} \ket{d}_{5} \ket{e}_{6}$~\cite{seki2020symmetry}.
The product $g$-state, however, belongs to the $q=0$ sector, and thus its time evolution under $\hat H(t)$ in Eq.~(\ref{eq:pvqe_ansatz}), which commutes with $\hat T$ (i.e., $[\hat T, \hat H(t)]=0$), cannot reach the subspace containing the target ground state with $q=\pi$.

To remedy this, we adopt a momentum-$\pi$ initial state defined as  
\begin{align}\label{eq: GHZ_state}
\ket{\psi_0} = \ket{\Psi_{q=\pi}} \equiv \frac{1}{\sqrt{2}} 
\left(|\uparrow \downarrow \uparrow \cdots \downarrow\rangle 
-|\downarrow \uparrow \downarrow \cdots \uparrow\rangle \right),
\end{align}
which satisfies ${\hat T} \ket{\Psi_{q=\pi}} 
= -\ket{\Psi_{q=\pi}}$.  
As shown in Figs.~\ref{fig4}(a) and \ref{fig4}(b), 
initializing the PVQE with this $q=\pi$ state drastically improves convergence for both $N=6$ and $N=10$, yielding significantly more accurate results. 
The corresponding optimized pulse profiles obtained at the final time-splitting step are presented in Figs.~\ref{fig4}(c) and \ref{fig4}(d).
The final optimized radii for the best-performing runs are $R=10.39$ $\upmu \text{m}$ for $N=6$ and $R=15.81$ $\upmu \text{m}$ for $N=10$, corresponding to nearest-neighbor Ising interactions of $J_{\rm nn}/h = 0.68$~MHz and $0.99$~MHz, respectively. The ensemble-averaged relative errors over 100 independent PVQE runs are $0.0051 \pm 0.0005$\% for $N=6$ and $1.6519 \pm 0.1677$\% for $N=10$.

Despite the promising results presented above, preparing the momentum $q=\pi$ state $\ket{\Psi_{q=\pi}}$ is not straightforward in optical tweezer arrays. 
In Appendix~\ref{app:state_preparation}, we propose an experimentally feasible method that exploits the fact that $\ket{\Psi_{q=\pi}}$ can be generated from the product $g$-state $\ket{g}^{\otimes N}$ using a Greenberger-Horne-Zeilinger (GHZ) type quantum circuit. 
Within analog quantum simulations using neutral atoms, the fundamental gates required for this GHZ circuit can be approximately realized through variational quantum gates, as discussed in Sec.~\ref{Sec:6}.
There, we also show that the variational quantum gate approach enables the experimental estimation of the expectation value of the Heisenberg Hamiltonian.

%%%%%%%%%%%%%%%%%%%%
%%%%%%%%%%%%%%%%%%%%
\begin{figure*} 
\includegraphics[width=\textwidth]{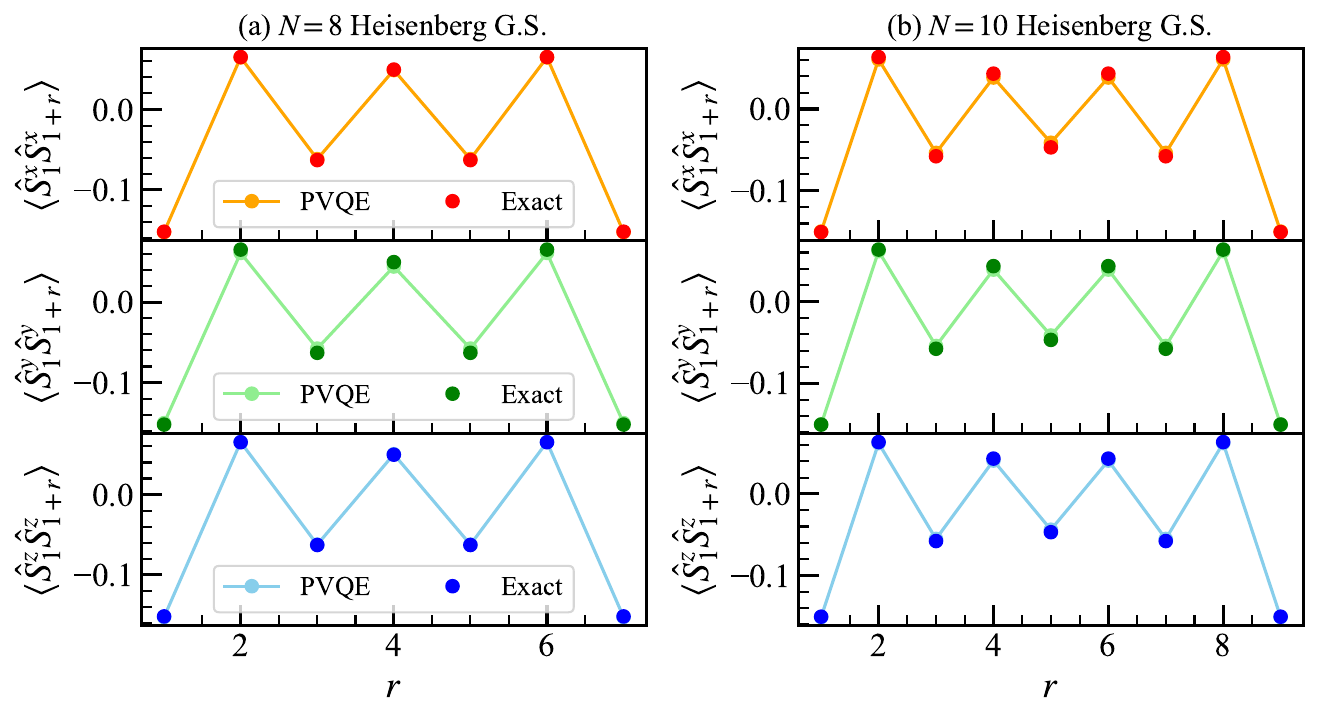}
\vspace{-6mm}
\caption{
Spin correlation functions of the optimized PVQE state for the one-dimensional antiferromagnetic Heisenberg model.  
(a) 
Results for $N=8$, corresponding to the best-performing run with the optimized pulse profiles shown in Fig.~\ref{fig3}(b).  
The red, green, and blue lines with circles represent the spin correlations $\langle S^x_i S^x_j \rangle$, $\langle S^y_i S^y_j \rangle$, and $\langle S^z_i S^z_j \rangle$, respectively.  
The PVQE results exhibit excellent agreement with those obtained from the numerically exact diagonalization method.  
(b) 
Same as panel (a), but for $N=10$.  
The corresponding optimized pulse profiles are shown in Fig.~\ref{fig4}(d).
}
\label{fig5}
\end{figure*}
%%%%%%%%%%%%%%%%%%%%
%%%%%%%%%%%%%%%%%%%%

The spin rotational symmetry inherent in the Heisenberg model, as described by the Hamiltonian $\hat H_{\rm XXX}$, manifests as isotropic spatial correlations in the ground state. 
Remarkably, the PVQE method not only reproduces the correct ground-state energy but also faithfully captures the SU(2)-symmetric spin correlations. 
Figure~\ref{fig5} shows that a well-converged PVQE state satisfies $\langle {\hat S}^{x}_{1}{\hat S}^{x}_{1+r} \rangle = \langle {\hat S}^{y}_{1}{\hat S}^{y}_{1+r} \rangle = \langle {\hat S}^{z}_{1}{\hat S}^{z}_{1+r} \rangle$ for all $r > 0$ in systems with $N=8$ and $N=10$. Here, $\langle\cdots\rangle$ denotes the expectation value with respect to the optimized PVQE state. 
These results indicate that the optimized state retains SU(2) symmetry, faithfully reflecting the symmetry of the target Hamiltonian, even though the variational ansatz defined in Eq.~(\ref{eq:pvqe_ansatz}) does not, in general, preserve SU(2) symmetry.
The PVQE results show quantitative agreement with numerically exact results obtained using the exact diagonalization method~\cite{weinberg2017quspin}.

%%%%%%%%%%%%%%%%%%%%%%%%%%%%%%%%%%%%%%%%%
%%%%%%%%%%%%%%%%%%%%%%%%%%%%%%%%%%%%%%%%%
\section{
PVQE for the mixed-field Ising model 
}
\label{Sec:5}
%%%%%%%%%%%%%%%%%%%%%%%%%%%%%%%%%%%%%%%%%
%%%%%%%%%%%%%%%%%%%%%%%%%%%%%%%%%%%%%%%%%

It is natural to consider extending the PVQE method to other generic interacting spin models that do not possess the full spin rotational symmetry of the Heisenberg model.
As a prototypical example, we focus on the one-dimensional mixed-field Ising (MFI) model with nearest-neighbor interactions~\cite{kitaev2010topological}. 
The Hamiltonian is given by  
\begin{align}\label{eq: MFI}
    {\hat H}_{\rm MFI} = J_{\rm I} \sum_{i=1}^{N}{\hat Z}_{i} {\hat Z}_{i+1} + \sum_{i=1}^{N}\left( h_x {\hat X}_{i} + h_z {\hat Z}_{i} \right), 
\end{align}
where $J_{\rm I} > 0$, $h_x$, and $h_z$ represent the antiferromagnetic Ising coupling, the transverse field, and the longitudinal field, respectively. Throughout this section, we assume periodic boundary conditions and set $J_{\rm I}=1$, so that all energy scales, including 
$h_x$ and $h_z$, are expressed in units of $J_{\rm I}$. 
When $h_z=0$, the Hamiltonian can be mapped to a free-fermion model via the Jordan-Wigner transformation~\cite{kitaev2010topological}, allowing an exact analytical treatment. 
In this limit, the model exhibits a second-order quantum phase transition at $h_x=1$. 
Beyond this exactly solvable point, however, the ground-state properties of Eq.~(\ref{eq: MFI}) must be investigated numerically. 
For $h_z \neq 0$, the low-energy spectrum is expected to host a variety of interesting features, including an emergent Ising field theory with $E_8$ symmetry that predicts a spectrum of massive quasiparticle branches~\cite{rutkevich2005large,coldea2010quantum,lamb2024ising,vovrosh2025meson}.

%%%%%%%%%%%%%%%%%%%%
%%%%%%%%%%%%%%%%%%%%
\begin{figure*} 
\includegraphics[width=\textwidth]{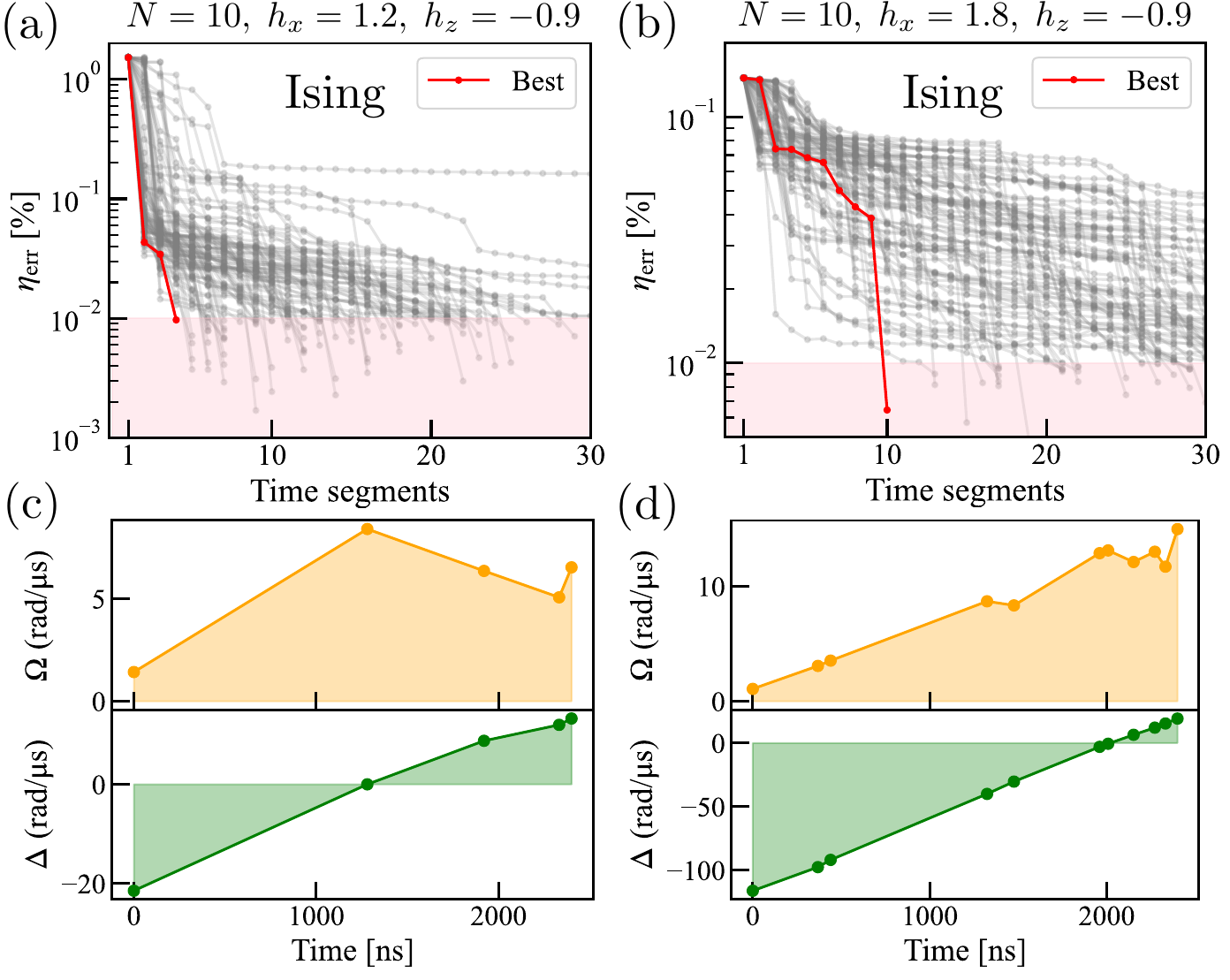}
\vspace{-6mm}
\caption{
Pulse-based VQE results for the one-dimensional MFI model with $N=10$ under periodic boundary conditions.  
(a) 
Relative error $\eta_{\rm err}$ tracked at each step of the time splitting process for 100 independent PVQE optimization runs. 
The MFI model parameters are set to $h_x = 1.2$ and $h_z = -0.9$, and the total pulse duration is fixed at $T=2.4$ $\upmu\text{s}$. 
The pink-shaded region indicates the target threshold $\epsilon = 0.01$: once  
$\eta_{\rm err}$ falls below this value, the time-splitting loop is terminated. 
The red line with circles highlights the best-performing run, which achieves convergence with the minimum number of segments. 
(b) 
Same as panel (a), but for $h_x = 1.8$.  
A total of 85 independent PVQE optimization runs are shown.  
(c) 
Optimized pulse profiles for the Rabi frequency $\Omega(t)$ and detuning $\Delta(t)$ at the final iteration, consisting of four linearly varying segments. 
These results correspond to the best-performing run highlighted in red in panel (a).  
(d) 
Same as panel (c), but for the best-performing run highlighted in red in panel (b), 
consisting of ten linearly varying segments.
}
\label{fig6}
\end{figure*}
%%%%%%%%%%%%%%%%%%%%
%%%%%%%%%%%%%%%%%%%%

Figure~\ref{fig6} presents PVQE results for the MFI model with $N=10$ and $h_z=-0.9$. 
Figures~\ref{fig6}(a) and \ref{fig6}(b) correspond to the transverse field values $h_x=1.2$ and $h_x=1.8$, respectively. For these settings, we perform 100 (for $h_x=1.2$) and 85 (for $h_x=1.8$) independent PVQE optimization runs. In both cases, nearly all runs successfully converge to the target ground state, achieving relative errors $\eta_{\rm err}$ below $0.1$\%. 
This favorable performance is not limited to these particular values of $h_x$, but is consistently observed across a broad range $0 < h_x \leq 2|h_z|$.
In all simulations, the time-splitting loops proceed up to a maximum of 30 segments, unless the relative error drops below the threshold $\epsilon=0.01$ at an earlier stage.
Notably, the best-performing runs achieve convergence within just four (for $h_x=1.2$) and ten (for $h_x=1.8$) time segments, as highlighted in red in Figs.~\ref{fig6}(a) and \ref{fig6}(b), respectively.
The corresponding optimized pulse profiles for the Rabi frequency $\Omega(t)$ and detuning $\Delta(t)$ at the final iteration are also shown in Figs.~\ref{fig6}(a) and \ref{fig6}(b) for $h_x=1.2$ and $h_x=1.8$, respectively.

\begin{table}[htbp]
\begin{ruledtabular}
\begin{tabular}{lcccc}
$h_x$ & mean $\eta_{\rm err}$ [\%] & mean $R$ [$\upmu$m] & $J_{\rm nn}(\langle R \rangle)$ [MHz]\\
\colrule
0.8 & 0.0103 $\pm$ 0.0006 &  13.6095 $\pm$ 0.0309 & 2.4362  \\
1.0 & 0.0150 $\pm$ 0.0012 &  13.2984 $\pm$ 0.0284 & 2.7987  \\
1.2 & 0.0093 $\pm$ 0.0016 &  14.7598 $\pm$ 0.0397 & 1.4972  \\
1.8 & 0.0156 $\pm$ 0.0010 &  13.9828 $\pm$ 0.0118 & 2.0710  \\
\end{tabular}
\end{ruledtabular}
\caption{
Mean relative errors of the ground-state energy and optimized final radii $\langle R\rangle$ for representative values of the transverse field $h_x$. The corresponding nearest-neighbor Ising interactions $J_{\rm nn}$, estimated at $\langle R\rangle$, are also listed in the fourth column. The longitudinal magnetic field is set to $h_z=-0.9$ and $N=10$.  
}
\label{tab1}
\end{table}

\begin{figure} 
\includegraphics[width=\columnwidth]{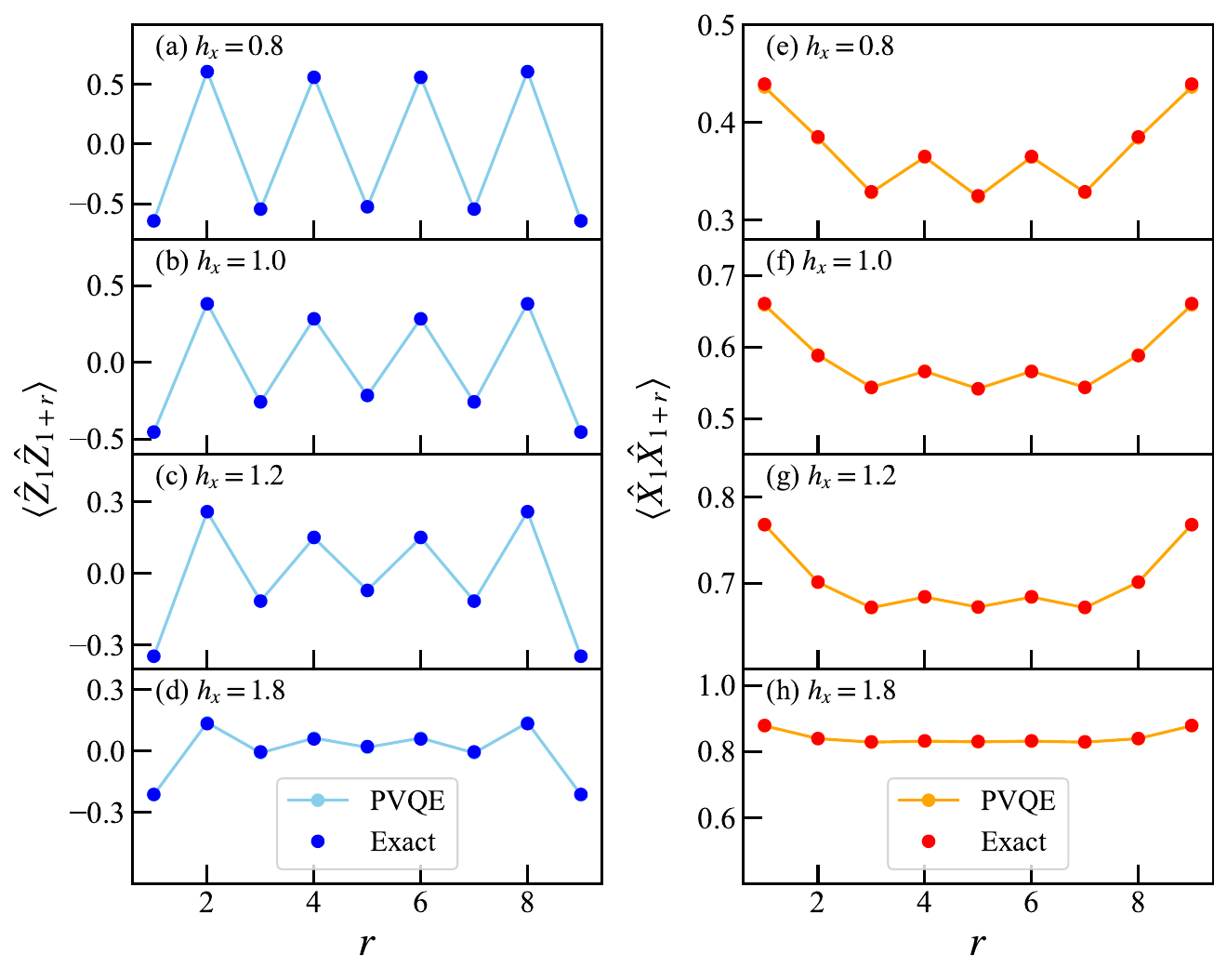}
\vspace{-6mm}
\caption{
Spin correlation functions in the optimized PVQE states for the one-dimensional MFI model with $h_z=-0.9$ and $N=10$ under periodic boundary conditions.
(a-d)
Longitudinal spin correlation functions $\langle {\hat Z}_{i}{\hat Z}_{j} \rangle$ evaluated from the best-performing PVQE optimization runs. 
Panels (a) through (d) correspond to $h_x=0.8$, $1.0$, $1.2$, and $1.8$, respectively. 
For comparison, the results obtained by the exact diagonalization method are also shown. 
(e-h)
Same as panels (a-d), but for the transverse spin correlation functions $\langle {\hat X}_{i} {\hat X}_{j} \rangle$. 
}
\label{fig7}
\end{figure}

Table~\ref{tab1} summarizes the averaged relative errors, as well as the averaged final radii $\langle R\rangle$ and the corresponding nearest-neighbor Ising interactions $J_{\rm nn}$ estimated at $\langle R\rangle$, computed over all runs that either satisfy the threshold condition or reach the maximum number of time segments. 
For representative values $h_x=0.8$, $1.0$, $1.2$, and $1.8$, the average relative error remains below $0.02$\%, demonstrating robust convergence of the PVQE method for the MFI model. 
These results indicate that, within the PVQE ansatz constructed from the time-evolution operator of the Ising-type Hamiltonian $\hat H(t)$ defined in Eq.~(\ref{eq:IsingHamiltonian}), the ground state of the MFI model is significantly more tractable than that of the Heisenberg model.  
Furthermore, as shown in Fig.~\ref{fig7}, the optimized PVQE states faithfully reproduce both the longitudinal and transverse spin correlation functions, $\langle \hat{Z}_i \hat{Z}_j \rangle$ and $\langle \hat{X}_i \hat{X}_j \rangle$, respectively, for representative values of $h_x$, in excellent agreement with the exact diagonalization results.

%%%%%%%%%%%%%%%%%%%%
%%%%%%%%%%%%%%%%%%%%
%%%%%%%%%%%%%%%%%%%%
%%%%%%%%%%%%%%%%%%%%

Finally, we note that the efficiency of the PVQE method for the MFI model can be attributed to the close resemblance between the target Hamiltonian and the Rydberg-atom Hamiltonian given in Eq.~\eqref{eq:IsingHamiltonian}. In particular, for the one-dimensional model with periodic boundary conditions considered here, the two Hamiltonians differ only by the presence of long-range interaction tail in the Rydberg setup. Consistent with this structural similarity, the optimized pulse profiles obtained via PVQE, such as those shown in Fig.~\ref{fig6}(d), resemble the quasi-adiabatic pulse schedules commonly used for ground-state preparation in Ising-type models~\cite{scholl2021quantum}.
This observation suggests that the global pulse optimization via PVQE may be directly applicable for preparing the many-body ground state of the analog system described by Eq.~(\ref{eq:IsingHamiltonian}). 
Such capability could be particularly useful for initiating nonequilibrium quantum quench dynamics in Rydberg atomic systems from a well-prepared ground state.

%%%%%%%%%%%%%%%%%%%%%%%%%%%%%%%%%%%%%%%%
%%%%%%%%%%%%%%%%%%%%%%%%%%%%%%%%%%%%%%%%
\section{
Measurement of the target Hamiltonian
}
\label{Sec:6}
%%%%%%%%%%%%%%%%%%%%%%%%%%%%%%%%%%%%%%%%
%%%%%%%%%%%%%%%%%%%%%%%%%%%%%%%%%%%%%%%%

In the preceding sections, we demonstrated the performance of the state-optimization component of the PVQE algorithm using two prototypical quantum spin models.  
In this section, we introduce a strategy to integrate the PVQE algorithm with the variational quantum gate approach~\cite{chevallier2024variational}, enabling experimental measurement of the target Hamiltonian.  
To illustrate the procedure, we focus on the Heisenberg model as a concrete example.
We note, however, that the same approach can be straightforwardly extended to the MFI model.

The cost function for the Heisenberg model defined in Eq.~(\ref{eq: AFH}) consists of the expectation values of local nearest-neighbor $XX$, $YY$, and $ZZ$ bond terms, and can be written as 
\begin{align}
    \langle {\hat H}_{\rm XXX} \rangle = \frac{J}{4} \sum_{j=1}^{N}\left ( \langle {\hat X}_{j} {\hat X}_{j+1} \rangle + \langle {\hat Y}_{j} {\hat Y}_{j+1} \rangle + \langle {\hat Z}_{j} {\hat Z}_{j+1} \rangle \right),
\end{align}
where $\langle \cdots \rangle$ denotes the expectation value with respect to the PVQE state defined in Eq.~(\ref{eq:pvqe_ansatz}). 
In typical Rydberg-atom experiments, the quantization axis is aligned along the $z$ direction. 
Consequently, $Z_i$ operators and their products, such as $Z_i Z_j Z_k \cdots$, are naturally accessible observables via fluorescence imaging measurements.

To evaluate bond correlations along the $x$ and $y$ directions, we apply global spin rotations that effectively rotate the quantization axis.  
The global $X$ and $Y$ rotation operators are defined as
\begin{align}
    {\hat R}_X(\theta) \equiv \bigotimes_{j=1}^{N} {\hat R}^{(j)}_x(\theta),\; 
    {\hat R}_Y(\theta) \equiv \bigotimes_{j=1}^{N} {\hat R}^{(j)}_y(\theta),
\end{align}
where $\theta \in {\mathbb R}$, ${\hat R}_\alpha(\theta) = \exp{-i\theta {\hat \sigma}_\alpha/2}$ denotes a single-qubit rotation around the $\alpha\,(= x,y)$ axis, with $\hat\sigma_\alpha$ being the corresponding Pauli operator.
These local rotation operators transform the ${\hat Z}_i$ operator into ${\hat X}_i$ or ${\hat Y}_i$ as follows: 
\begin{align*}
R_y(\theta) {\hat Z}_i R_y(-\theta) 
&= \cos \theta {\hat Z}_i + \sin \theta {\hat X}_i, \\
R_x(\theta) {\hat Z}_i R_x(-\theta) 
&= \cos \theta {\hat Z}_i - \sin \theta {\hat Y}_i. 
\end{align*}
As is well known, the bond correlators $\langle {\hat X}_i {\hat X}_j \rangle$ and $\langle {\hat Y}_i {\hat Y}_j \rangle$ can thus be evaluated by measuring the operator $ {\hat Z}_i {\hat Z}_j $ after applying suitable global rotations:
\begin{align}
    \langle {\hat X}_i {\hat X}_j \rangle 
    &= \bra{\psi_0} {\hat U}_{\rm Ising}^{\dagger} {\hat R}_Y(-\pi/2) {\hat Z}_i {\hat Z}_j {\hat R}_Y(\pi/2) {\hat U}_{\rm Ising} \ket{\psi_0}, \nonumber \\
    \langle {\hat Y}_i {\hat Y}_j \rangle 
    &= \bra{\psi_0} {\hat U}_{\rm Ising}^{\dagger} {\hat R}_X(-\pi/2) {\hat Z}_i {\hat Z}_j {\hat R}_X(\pi/2) {\hat U}_{\rm Ising} \ket{\psi_0}, 
\end{align}
where $\hat U_{\rm Ising}$ is the time-evolution operator given in Eq.~(\ref{eq:Uising}). 
Note that the sign of the $\pi/2$ rotation angle does not affect the resulting expectation values.
In analog quantum simulations, such global rotations can be effectively implemented using the variational quantum gate approach~\cite{chevallier2024variational}, 
in which both single- and two-qubit gates are replaced by globally optimized analog pulse sequences. 
Based on results from Ref.~\cite{chevallier2024variational}, for $J_{\rm nn}/h \approx 2.3$~MHz in a circular geometry, 
the pulse duration required to realize a global $\pi/2$-rotation is approximately $0.9$ $\upmu\text{s}$, well below the typical coherence time of $\sim6$ $\upmu\text{s}$~\cite{scholl2021quantum}.

%%%%%%%%%%%%%%%%%%%%
%%%%%%%%%%%%%%%%%%%%
\begin{figure} 
\includegraphics[width=\columnwidth]{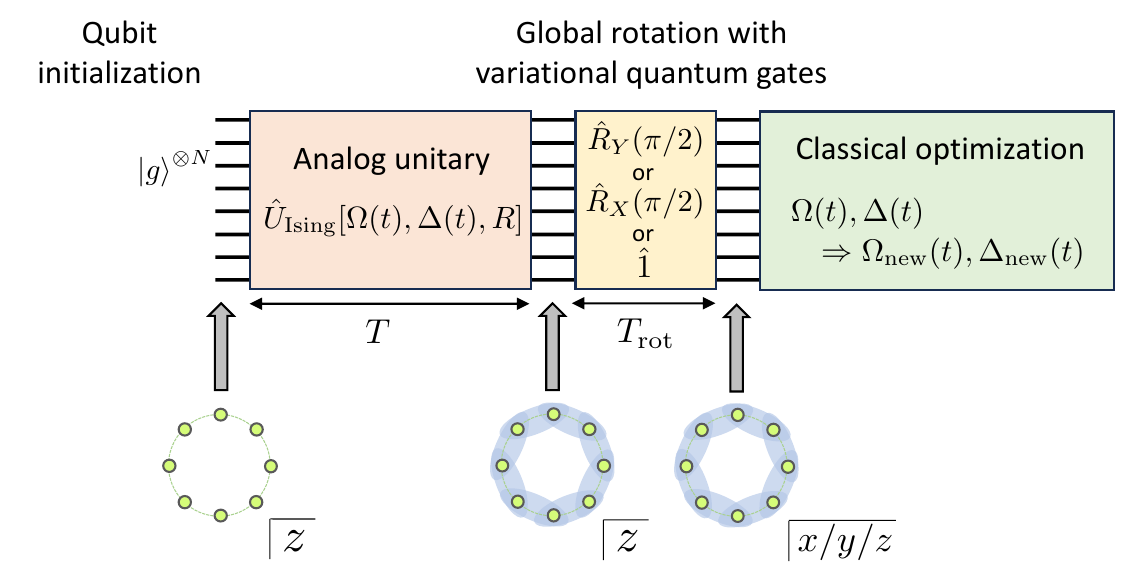}
\vspace{-6mm}
\caption{
Pulse-based VQE combined with the variational quantum gate method for $N=8$.
The atoms are first initialized in the product state $\ket{g \cdots g}$ and then evolved under the analog unitary ${\hat U}_{\rm Ising}(T)$ generated by the time dependent Hamiltonian $\hat H(t)$ with $\Omega(t)$ and $\Delta(t)$.
After this evolution, a global rotation operator $\hat U_{\rm rot} (T_{\rm rot})$, implementing a basis change to the $x$, $y$, or $z$ direction, is applied to enable measurement of the desired observables. 
To evaluate the cost function via classical optimization, the expectation values of bond operators are estimated by repeating the same experimental sequence. 
Once a new set of parameters, $\Omega_{\rm new}(t)$ and $\Delta_{\rm new}(t)$, is obtained, the next iteration of the PVQE loop is performed. 
}
\label{fig8}
\end{figure}
%%%%%%%%%%%%%%%%%%%%
%%%%%%%%%%%%%%%%%%%%

As illustrated in Fig.~\ref{fig8}, the PVQE state including the global rotation is given by 
\begin{align}\label{eq: PVQE_Rotation}
    \ket{\psi(T_{*})} = {\hat U}_{\rm rot}(T_{\rm rot}) {\hat U}_{\rm Ising}(T) \ket{\psi_{0}}, 
\end{align}
where $T_{*} =T + T_{\rm rot}$ is the total pulse duration, and $T_{\rm rot}$ is the time required to implement the global rotation.
In Ref.~\cite{chevallier2024variational}, the unitary operator ${\hat U}_{\rm rot}(T_{\rm rot})$ representing global rotation is obtained by optimizing analog pulse sequences for a fixed atomic geometry corresponding to a nearest-neighbor Ising interaction of $J_{\rm nn}/h \approx 2.3$~MHz.
To leverage these results and assess whether the total duration $T_{*}$ remains within the coherence time, we perform a modified PVQE optimization for ${\hat U}_{\rm Ising}(T)$ in which the array radius is fixed at $R_* = 11.1$~$\upmu\text{m}$, matching the same interaction strength.
Unlike the variable-$R$ optimization in Sec.~\ref{Sec:4}, we employ the gradient-based L-BFGS-B algorithm (from the SciPy library) instead of the Nelder-Mead method. 
The pulse duration is set to $T = 2.4$ $\upmu\text{s}$, which, as shown previously, is sufficient to accurately prepare the ground state.

\begin{figure} 
\includegraphics[width=\columnwidth]{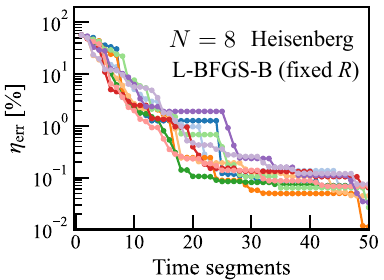}
\vspace{0mm}
\caption{
Fixed-radius pulse-based VQE for the one-dimensional antiferromagnetic Heisenberg model with $N=8$ under periodic boundary conditions. 
Relative error $\eta_{\rm err}$ is shown at each step of the time-splitting process for 10 representative runs that achieved good convergence, out of 100 independent PVQE optimizations. 
The best-performing result, shown in orange, reaches a relative error on the order of $10^{-2}$\% after 50 time segments. 
The total pulse duration is fixed at $T=2.4$ $\upmu\text{s}$. 
}
\label{fig9}
\end{figure}

Figure~\ref{fig9} confirms that the PVQE procedure with fixed $R$ and gradient-based optimization also yields highly accurate approximations to the target ground state. 
In this setup, the total duration $T_* \approx 3.3$ $\upmu\text{s}$ remains well below typical coherence times (on the order of $6\ \upmu\text{s}$), indicating that the protocol in Eq.~(\ref{eq: PVQE_Rotation}) is both experimentally feasible and implementable on near-term quantum devices such as those developed by Pasqal~\cite{silverio2022pulser} and QuEra~\cite{wurtz2023aquila}. 
Moreover, we expect that this hybrid approach, combining PVQE with pulse-level gate optimization, is not restricted to fixed-radius configurations and can be extended to setups with variable $R$.  
A detailed investigation of such extensions is left for future work.

%%%%%%%%%%%%%%%%%%%%..
%%%%%%%%%%%%%%%%%%%%
%%%%%%%%%%%%%%%%%%%%
%%%%%%%%%%%%%%%%%%%%

%%%%%%%%%%%%%%%%%%%%%%%%%
%%%%%%%%%%%%%%%%%%%%%%%%%
\section{
Conclusions
}
\label{Sec:7}
%%%%%%%%%%%%%%%%%%%%%%%%%
%%%%%%%%%%%%%%%%%%%%%%%%%

In summary, we have investigated the performance of the PVQE implemented using Rydberg atoms in optical tweezer arrays. 
We demonstrated that the variational quantum algorithm, incorporating random pulse segmentation, can achieve high accuracy in approximating the many-body ground states of the one-dimensional antiferromangnetic Heisenberg and MFI models for systems of up to ten qubits. 
The optimized variational states not only reproduce the ground-state energies with good precision but also capture the corresponding spin correlation functions faithfully. 
Furthermore, we discussed how the variational quantum gate approach, an analog realization of digital quantum gates, can be utilized to facilitate the measurement of target many-body Hamiltonians, indicating that direct experimental validation of our results is feasible with current quantum hardware.

In future studies, we plan to investigate the scalability of this digital-analog variational quantum algorithm using state-of-the-art numerical techniques such as tensor network methods~\cite{xiang2023density}. 
Furthermore, the PulserDiff framework~\cite{abramavicius2025pulserdiff}, which enables gradient-based optimization of pulse sequences, may offer enhanced control fidelity and improved convergence, potentially overcoming limitations associated with our current pulse duration scheme.

{\it Note added---}
While finalizing this manuscript, we became aware of a recent work~\cite{singh2025ground} that also explores pulse-level VQE for the Heisenberg and Ising models using Rydberg atom platforms. 
While their study focuses primarily on the mathematical structure of variational optimization using Lie algebras and piecewise-constant pulse sequences, our approach emphasizes experimentally realistic pulse schedules based on linear ramps and introduces a practical scheme for measuring target Hamiltonians.

%%%%%%%%%%%%%%%%%%%%%%%%%%%%%
%%%%%%%%%%%%%%%%%%%%%%%%%%%%%
\begin{acknowledgments}

We thank Takafumi Tomita, Masaya Kunimi, Kazuhiro Seki, and Tomonori Shirakawa for valuable discussions. 
Numerical simulations based on Pulser~\cite{silverio2022pulser} were performed on the HOKUSAI supercomputing system at RIKEN. 
This work was partially supported by Project No.~JPNP20017, funded by the New Energy and Industrial Technology Development Organization (NEDO), Japan. 
We acknowledge support from JSPS KAKENHI (Grant Nos.~JP21H04446 and JP25K17321)
from the Ministry of Education, Culture, Sports, Science and Technology (MEXT), Japan.
Additional funding was provided by 
JST COI-NEXT (Grant No.~JPMJPF2221) and by 
the Program for Promoting Research of the Supercomputer Fugaku (Grant No.~MXP1020230411) from MEXT, Japan.  
We further acknowledge support from 
the RIKEN TRIP initiative (RIKEN Quantum) and
the COE research grant in computational science from Hyogo Prefecture and Kobe City through the Foundation for Computational Science. 
Exact diagonalization calculations were performed using QuSpin~\cite{weinberg2017quspin}. 
Pasqal's team acknowledges funding from the European Union through the project PASQuanS2.1 (HORIZON-CL4-2022-QUANTUM02-SGA, Grant Agreement 101113690).

\end{acknowledgments}
%%%%%%%%%%%%%%%%%%%%%%%%%%%%%
%%%%%%%%%%%%%%%%%%%%%%%%%%%%%

\appendix

%%%%%%%%%%%%%%%%%%%%%%%%%%%
%%%%%%%%%%%%%%%%%%%%%%%%%%%
\section{
Number of Nelder-Mead iterations for the Heisenberg model
}
\label{app:iterations}
%%%%%%%%%%%%%%%%%%%%%%%%%%%
%%%%%%%%%%%%%%%%%%%%%%%%%%%

%%%%%%%%%%%%%%%%%%%%
%%%%%%%%%%%%%%%%%%%%
\begin{figure*} 
\includegraphics[width=\textwidth]{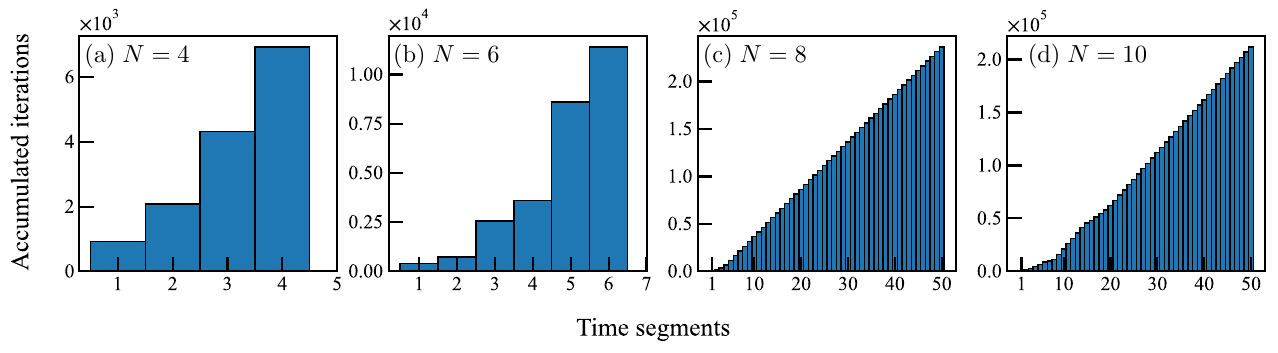}
\vspace{-6mm}
\caption{ 
Total number of Nelder-Mead iterations accumulated up to each time segment to obtain the best-performing PVQE results for the antiferromagnetic Heisenberg model with $N=4$, 6, 8, and 10 under periodic boundary conditions. 
}
\label{fig10}
\end{figure*}
%%%%%%%%%%%%%%%%%%%%
%%%%%%%%%%%%%%%%%%%%

Figure~\ref{fig10} shows the number of Nelder-Mead iterations required to obtain the best-performing results for the Heisenberg model. 
For example, in the case of $N=4$ [Fig.~\ref{fig10}(a)], a total of 6935 Nelder-Mead iterations were performed by the final step of the time-splitting process. 
The number of iterations increases with the number of time segments introduced through successive splitting. 
For $N=8$ and $N=10$, which require 50 time segments to achieve a relative error on the order of $10^{-1}$\%, the total number of iterations amounts to 236495 and 211965, as shown in Figs.~\ref{fig10}(c) and \ref{fig10}(d), respectively.

%%%%%%%%%%%%%%%%%%%%%%%%%%%
%%%%%%%%%%%%%%%%%%%%%%%%%%%
\section{
Experimental implementation of the $q=\pi$ state $|\Psi_{q=\pi}\rangle$
}
\label{app:state_preparation}
%%%%%%%%%%%%%%%%%%%%%%%%%%%
%%%%%%%%%%%%%%%%%%%%%%%%%%%

We present concise protocols for dynamically preparing the $q=\pi$ state $|\Psi_{q=\pi}\rangle$ for arbitrary $N$ in analog quantum simulations using optically confined neutral atoms.
Two complementary approaches are described: 
(i) a quantum circuit approach, in which local quantum gates are assumed to be implemented via optimized pulse controls, and 
(ii) a Hamiltonian-based approach, which utilizes an $N$-body interaction Hamiltonian combined with a layer of single-site gates.

%%%%%%%%%%%%%%%%%%%%
%%%%%%%%%%%%%%%%%%%%
\begin{figure} 
$$
\Qcircuit @C=1em @R=.7em {
  \lstick{\ket{g}_1} & \gate{\rm H} & \ctrl{1} & \qw      & \qw      & \qw      & \qw      & \qw & \qw  \\
  \lstick{\ket{g}_2} & \qw          & \targ    & \ctrl{1} & \qw      & \qw      & \qw      & \qw & \qw  \\
  \lstick{\ket{g}_3} & \qw          & \qw      & \targ    & \ctrl{1} & \qw      & \qw      & \qw & \qw  \\
  \lstick{\ket{g}_4} & \qw          & \qw      & \qw      & \targ    & \ctrl{1} & \qw      & \qw & \qw  \\
  \lstick{\ket{g}_5} & \qw          & \qw      & \qw      & \qw      & \targ    & \ctrl{1} & \qw & \qw  \\
  \lstick{\ket{g}_6} & \qw          & \qw      & \qw      & \qw      & \qw      & \targ    & \qw & \qw 
}
$$
$$
\Qcircuit @C=1em @R=.7em {
  \lstick{\ket{g}_1} & \gate{\rm H} & \ctrl{3} & \ctrl{1}  & \qw      & \qw & \qw  \\
  \lstick{\ket{g}_2} & \qw          & \qw      & \targ     & \ctrl{1} & \qw & \qw  \\
  \lstick{\ket{g}_3} & \qw          & \qw      & \qw       & \targ    & \qw & \qw  \\
  \lstick{\ket{g}_4} & \qw          & \targ    & \ctrl{1}  & \qw      & \qw & \qw  \\
  \lstick{\ket{g}_5} & \qw          & \qw      & \targ     & \ctrl{1} & \qw & \qw  \\
  \lstick{\ket{g}_6} & \qw          & \qw      & \qw       & \targ    & \qw & \qw  
}
$$
\vspace{0mm}
\caption{
(Top) 
Quantum circuit representation of the GHZ unitary operator used to prepare the $q=\pi$ state 
$|\Psi_{q=\pi}\rangle$ in Eq.~(\ref{eq:ghz}). 
(Bottom)
Optimized version of the GHZ circuit with a logarithmic depth in the number of qubits 
}
\label{fig11}
\end{figure}
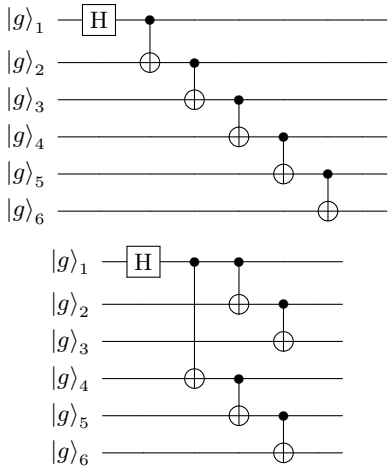
%%%%%%%%%%%%%%%%%%%%
%%%%%%%%%%%%%%%%%%%%

For later use, 
we define the Hadamard gate in terms of the $|g\rangle$-$|r\rangle$ basis as 
\begin{align}
{\hat {\rm H}} = \frac{1}{\sqrt{2}}\left( \ket{r}\bra{r} - \ket{g}\bra{g} + \ket{r}\bra{g} + \ket{g}\bra{r} \right).
\end{align}
This gate transforms the basis states into equal-weighted superpositions as 
\begin{align}
{\hat {\rm H}} \ket{g} = \frac{1}{\sqrt{2}}\ket{r} - \frac{1}{\sqrt{2}}\ket{g},\; \; 
{\hat {\rm H}} \ket{r} = \frac{1}{\sqrt{2}}\ket{r} + \frac{1}{\sqrt{2}}\ket{g}. 
\end{align}
We also define the controlled-NOT (CNOT) gate acting on a controlled qubit $q_c$ and a target qubit $q_t$ as 
\begin{align}
{\widehat {\rm CNOT}}_{q_c, q_t} = \ket{r}\bra{r}_{q_c} \otimes {\hat I}_{q_t} + \ket{g}\bra{g}_{q_c} \otimes {\hat X}_{q_t}.
\end{align}
These definitions are consistent with the standard conventions in quantum information theory~\cite{nielsen2010quantum}, under the correspondence $\ket{g} \leftrightarrow \ket{1}$ and $\ket{r} \leftrightarrow \ket{0}$.

Given the above notational setup, we now define a Greenberger-Horne-Zeilinger (GHZ) circuit operator as ${\widehat C}_{\rm GHZ} = \left( \prod_{j=1}^{N-1}{\widehat {\rm CNOT}}_{j,j+1} \right) {\widehat {\rm H}_1}$, which generates the GHZ state by acting on the fully-up product state $\ket{r \cdots r}=\ket{0 \cdots 0}$ as ${\widehat C}_{\rm GHZ} \ket{r \cdots r} = \frac{1}{\sqrt{2}}(\ket{r\cdots r} + \ket{g \cdots g})$~\cite{facchi2011greenberger}. 
We observe that this global operator ${\widehat C}_{\rm GHZ}$ also transforms the fully-down product state $\ket{g \cdots g}$ into the desired $q=\pi$ state $|\Psi_{q=\pi}\rangle$, i.e.,   
\begin{align} 
    \ket{\Psi_{q=\pi}} = {\widehat C}_{\rm GHZ} \ket{g \cdots g}.
    \label{eq:ghz}
\end{align}
Indeed, this transformation proceeds as follows: 
\begin{align*}
\ket{g\cdots g} 
&\xrightarrow{ {\rm H}_1 } \frac{1}{\sqrt{2}}\ket{rg \cdots g} - \frac{1}{\sqrt{2}}\ket{g \cdots g}, \\
&\xrightarrow{ {\rm CNOT}_{1,2}} \frac{1}{\sqrt{2}}\ket{rg \cdots g} - \frac{1}{\sqrt{2}}\ket{grg \cdots g},  \\
&\xrightarrow{ {\rm CNOT}_{2,3}} \frac{1}{\sqrt{2}}\ket{rgrg \cdots g} - \frac{1}{\sqrt{2}}\ket{grg \cdots g},  \\
&\xrightarrow{ {\rm CNOT}_{3,4}} \frac{1}{\sqrt{2}}\ket{rgrg \cdots g} - \frac{1}{\sqrt{2}}\ket{grgrg \cdots g}, \\
& \;\;\; \cdots \\
&\xrightarrow{ {\rm CNOT}_{N-1,N}} \frac{1}{\sqrt{2}}\ket{rgrgrg} - \frac{1}{\sqrt{2}}\ket{grgrgr}. 
\end{align*}

A schematic diagram of the GHZ circuit for $N=6$ is shown in the top panel of Fig.~\ref{fig11}.  
In analog quantum simulations, digital quantum gates can be effectively implemented via variational quantum gates, as discussed in Sec.~\ref{Sec:6}.  
However, for a nearest-neighbor interaction of $J_{\rm nn}/h \sim 2.3$ MHz, a single CNOT gate requires approximately $6.2$ $\upmu\text{s}$~\cite{chevallier2024variational}.  
Therefore, the practical implementation of the full GHZ circuit may be limited by the short coherence time of current quantum devices.  
Nonetheless, this limitation is expected to be alleviated with continued advances in experimental technologies.
Furthermore, we note that the GHZ circuit described above can be transformed into an optimized version with a logarithmic circuit depth in the number of qubits, as illustrated in the bottom panel of Fig.~\ref{fig11}~\cite{mooney2021generation, de2024quantum}.
Such optimization may allow for the preparation of the $q = \pi$ state with significantly shorter pulse durations compared to the naive GHZ circuit.

As an alternative to the quantum circuit approach described above, one can consider a Hamiltonian evolution scheme. 
Specifically, we introduce a string-type unitary operator parameterized by an angle $\theta$~\cite{facchi2011greenberger}, defined as 
\begin{align}
    {\hat U}_{\rm flip}(\theta) = \exp[-i \frac{\theta}{2} {\hat Y}_{1} {\hat X}_{2} \cdots {\hat X}_{N}].
\end{align}
This operator generates a continuous-time unitary evolution governed by the Pauli string $\hat{Y}_1 \hat{X}_2 \cdots \hat{X}_N$, with $\theta$ controlling the evolution duration.  
In particular, for $\theta = \pi/2$, 
the squaring properties of Pauli strings lead to 
\begin{align}
    {\hat U}_{\rm flip}(\pi/2) = \frac{1}{\sqrt{2}}{\hat 1} - \frac{i}{\sqrt{2}}{\hat Y}_{1} {\hat X}_{2} \cdots {\hat X}_{N}. 
\end{align}
Using this operator, the following transformation can be implemented:
\begin{align*}
    \ket{gg \cdots gg}
    &\xrightarrow{{\hat X}_2 {\hat X}_4 \cdots } 
    \ket{gr \cdots gr}, \\
    &\xrightarrow{ {\hat U}_{\rm flip}(\pi/2) } 
    \frac{1}{\sqrt{2}} \ket{gr \cdots gr} - \frac{1}{\sqrt{2}} \ket{rg \cdots rg}.
\end{align*}
We assume that such an $N$-body unitary operator could, in principle, be realized via optimized pulse control techniques in analog quantum simulations.

% Bibtex 
\bibliography{ref}

\end{document}